\documentclass[prd,aps,showpacs,tightenlines,nofootinbib,preprint,preprintnumbers]{revtex4}
\usepackage{graphicx}
\usepackage{amsfonts}
\usepackage{amssymb}
\usepackage{latexsym}
\usepackage{cancel}
\usepackage{color}
\input{colordvi.tex}

 \newcommand{\CL}{{\cal L}}
 
\newcommand{\CO}{{\cal O}}

\newcommand{\bear}{\begin{array}}  \newcommand{\eear}{\end{array}}
\newcommand{\bea}{\begin{eqnarray}}  \newcommand{\eea}{\end{eqnarray}}
\newcommand{\beq}{\begin{equation}}  \newcommand{\eeq}{\end{equation}}
\newcommand{\bef}{\begin{figure}}  \newcommand{\eef}{\end{figure}}
\newcommand{\bec}{\begin{center}}  \newcommand{\eec}{\end{center}}
\newcommand{\non}{\nonumber}  
\newcommand{\lmk}{\left(}  \newcommand{\rmk}{\right)}
\newcommand{\lkk}{\left[}  \newcommand{\rkk}{\right]}
  
  \newcommand{\abs}[1]{\vert{#1}\vert}

\newcommand{\bib}{\bibitem} 
\newcommand{\la}{\left\langle} \newcommand{\ra}{\right\rangle}

\newcommand{\Sla}[1]{{\not \! \! {#1}}}

\newcommand{\slaD}{\Sla{D}}

\def\AAA#1#2#3{Astron. Astrophys. {\bf #1}, #2 (20#3)}

\def\EPLL#1#2#3{Europhys. Lett. {\bf #1}, #2 (20#3)}
\def\GRG#1#2#3{Gen. Relativ. Gravit. {\bf #1}, #2 (19#3)}

\def\JCAPP#1#2#3{J. Cosmol. Astropart. Phys. {\bf #1}, #2 (20#3)}

\def\JHEPP#1#2#3{J. High Energy Phys. {\bf #1}, #2 (20#3)}

\def\MNRASS#1#2#3{Mon. Not. R. Astron. Soc. {\bf #1}, #2 (20#3)}

\def\NATT#1#2#3{Nature (London) {\bf #1}, #2 (20#3)}
\def\NPB#1#2#3{Nucl. Phys. {\bf B#1}, #2 (19#3)}
\def\NPBB#1#2#3{Nucl. Phys. {\bf B#1}, #2 (20#3)}

\def\PLBB#1#2#3{Phys. Lett. B {\bf #1}, #2 (20#3)}
\def\PLBold#1#2#3{Phys. Lett. {\bf#1B}, #2 (19#3)}

\def\PRD#1#2#3{Phys. Rev. D {\bf #1}, #2 (19#3)}
\def\PRDD#1#2#3{Phys. Rev. D {\bf #1}, #2 (20#3)}

\def\PRL#1#2#3{Phys. Rev. Lett. {\bf#1}, #2 (19#3)}
\def\PRLL#1#2#3{Phys. Rev. Lett. {\bf#1}, #2 (20#3)}

\def\RMPP#1#2#3{Rev. Mod. Phys. {\bf #1}, #2 (20#3)}

\begin{document}


\title{Time variation of proton-electron mass ratio and fine structure
constant with runaway dilaton}
\author{Takeshi Chiba}
\affiliation{ Department of Physics, College of Humanities and Sciences,
Nihon University, Tokyo 156-8550, Japan}
\author{Tatsuo Kobayashi}
\affiliation{Department of physics, Kyoto University, Kyoto 606-8502,
Japan}
\author{Masahide Yamaguchi}
\affiliation{Department of Physics and Mathematics, Aoyama Gakuin
University, Sagamihara 229-8558, Japan}
\author{Jun'ichi Yokoyama} 
\affiliation{Research Center for the Early Universe (RESCEU), Graduate
School of Science, The University of Tokyo, Tokyo, 113-0033, Japan}

\date{\today}
\preprint{KUNS-2041 \cr RESCEU-25/06 }
%

\begin{abstract}
Recent astrophysical observations indicate that the proton-electron mass
ratio and the fine structure constant have gone through nontrivial time
evolution.  We discuss their time variation in the context of a dilaton
runaway scenario with gauge coupling unification at the string scale
$M_{\rm s}$. We show that the choice of adjustable parameters allows
them to fit the same order magnitude of both variations and their
(opposite) signs in such a scenario.
\end{abstract}


\maketitle


\section{Introduction}

In unified theories of fundamental interactions, a variety of
fundamental constants are not necessarily ``constant'' but can vary as a
function of spacetime. Therefore, many experiments and observations have
been done to test the constancy of various fundamental constants
\cite{uzan}.

Among them, several groups report nonvanishing
 time variation of some of the fundamental
constants. For example, Murphy {\it et al.} report time variability
of the fine structure constant $\alpha$ by use of absorption systems in
the spectra of distant quasars \cite{webb}. They found that the fine
structure constant $\alpha$ was smaller in the past,
\beq
  \frac{\Delta \alpha}{\alpha} = (-0.543\pm 0.116)\times 10^{-5},  
  \label{eq:dalpha}
\eeq
for the redshift range $0.2 < z < 3.7$,\footnote{In \cite{webb2}, the
data is updated but the time variation has almost the same value $\Delta
\alpha / \alpha = (-0.57\pm 0.11)\times 10^{-5}$.} though similar
observations of other groups do not necessarily reproduce this result
\cite{SCP,LCM}. The linear interpolation of such a change yields the
rate of the change,\footnote{In this paper, we use the units of
$c=\hbar=1$.}
\beq
  \frac{\dot{\alpha}}{\alpha} = \CO(10^{-15})
 \,\,\, {\rm yr^{-1}} = \CO(10^{-65}) M_{G},  \label{alphayear}
\eeq
where the dot represents the time derivative and $M_{G}=1/\sqrt{8\pi G}$
is the reduced Planck scale.

The observations of ${\rm H}_2$ spectral lines in the Q 0347-383 and Q
0405-443 quasars also suggest a fractional change in the proton-electron
mass ratio $\mu = m_{\rm p} / m_{\rm e}$,
\beq
  \frac{\Delta \mu}{\mu} = (2.4 \pm 0.6) \times 10^{-5} ,
  \label{eq:dmu}
\eeq
for a weighted fit \cite{mpme}, which implies that it has decreased over
the last 12 Gyr. The linear interpolation of such a change yields the
rate of the change,
\beq
  \frac{\dot{\mu}}{\mu} \simeq - 2.0 \times 10^{-15}
 \,\,\, {\rm yr^{-1}} = -1.7\times 10^{-65}M_{G}.  \label{muyear}
\eeq
Thus, though
there are still large uncertainties, the hints of the time variation of
fundamental constants are found.
 
On the other hand, from the theoretical point of view, it is natural to
allow time and space dependence of fundamental constants. In fact, 
superstring theory, which is expected to unify all fundamental
interactions, predicts the existence of a scalar partner $\phi$ (called
dilaton) of the tensor graviton, whose expectation value determines the
string coupling constant $g_s = e^{\phi / 2}$ \cite{witten}. The
couplings of the dilaton to matter induces the violation of the equivalence
principle and hence generates deviations from general
relativity. Therefore, though the dilaton is predicted to be massless at
tree level, it is usually assumed that it acquires a sufficiently large
mass, associated with supersymmetry breaking, to satisfy the present
experimental constraints on the equivalence principle.
 
However, Damour and Polyakov proposed another possibility which can
naturally reconcile a massless dilaton with experimental constraints
\cite{DP}. They pointed out that full string-loop effects modify the
four dimensional effective low-energy action as
\beq
  S = \int d^4 x \sqrt{\widetilde{g}} 
    \lmk \frac{B_g (\phi)}{\alpha'} \, \widetilde{R} 
           + \frac{B_{\phi} (\phi)}{\alpha'} \, 
           \lbrack 2 \widetilde{\Box_{}^{}} \phi -
           (\widetilde{\nabla} \phi)^2 \rbrack 
           - \frac{1}{4} \, B_F(\phi) \, \widetilde{F}^2 - V 
           + \cdots 
    \rmk\,.
\label{eq:action}
\eeq
Here $B_i(\phi)$ ($i = g, \phi, F, \ldots$) are $\phi$ dependent
coupling functions. String scale $M_{\rm s}$ is given by $M_{\rm s} =
\alpha^{'-1/2}$. 
In the weak coupling limit $g_s \rightarrow 0$
($\phi \rightarrow - \infty$), they are expanded into
\beq
  B_i (\phi) = e^{-\phi} + c_0^{(i)} + c_1^{(i)}\,e^{\phi} 
           + c_2^{(i)}\,e^{2\phi} + \cdots,
\eeq
which comes from genus expansion of string theory with $B_i = \Sigma_n
\, g_s^{2(n-1)} c_n^{(i)}$ $(n = 0,1,2,\ldots)$.  Assuming a universality
of the dilaton coupling functions, that is, the existence of a value
$\phi_m$ of $\phi$ which extremizes all the coupling functions $B_i^{-1}
(\phi)$, it has been shown that, during a primordial inflationary stage,
the dilaton evolves towards the special value $\phi_m$, at which it
decouples from matter (so-called ``Least Coupling Principle'')
\cite{DP}. Subsequent (slight) change of the dilaton induces the time
variation of fundamental constants. Note that the dilaton becomes almost
homogeneous in space during inflation so that spatial variations of
fundamental constants are expected to be much smaller than their time
variations.

On the other hand, in the infinite bare coupling limit $g_s \rightarrow
\infty$ ($\phi \rightarrow + \infty$), it is suggested that all the coupling
functions have smooth finite limits \cite{IBC},
\beq
  B_i (\phi) = C_i + A_i e^{-\phi} = C_i (1 + d_i e^{-\phi}),
  \label{eq:cf}
\eeq
with $d_i \equiv A_i/C_i$. Since $A_i$ ($d_i$) is expected to be
positive in the ``large N''-type toy model of \cite{IBC}, we assume that
all $A_i$'s are positive in this paper. All the coupling functions are
extremized (minimized) at $\phi_m = + \infty$. In this case, also,
the dilaton evolves towards its fixed point at infinity during inflation so
that it decouples from matter \cite{GPV,DPV}. In Ref. \cite{DPV}, 
assuming the dilaton coupling to dark matter and/or
dark energy, the magnitude of the time variation of the fine structure 
constant
is estimated.

In this paper, we discuss time variations of the 
proton-electron mass ratio
and the fine structure constant in the context of a dilaton 
runaway scenario
with gauge coupling unification at the string scale $M_{\rm s}$. 
We show that our model can account for the putative time variation of 
these constants of the same magnitude with the opposite
 signs.

The rest of the paper is organized as follows.  In the next section, we
calculate time variation of the proton-electron mass ratio and that of
the fine structure constant by taking into account effects associated
with thresholds in renormalization group running and variations in the
vacuum expectation value (VEV) of the Higgs field.\footnote{For related
works in the context of GUT, see \cite{varying}.}  We then apply it to
the specific case of a runaway dilaton scenario in \S III.  In \S IV we
show that our model can not only fit the observations well but also is
testable by the experiments to verify the equivalence principle.
Finally \S V is devoted to the conclusion.

\section{Renormalization group analysis of the time variation of
the fundamental constants}

In this section, we calculate time variation rates of fundamental
constants such as
the proton-electron mass ratio
and the fine structure constant from a fundamental point of
view using renormalization group analysis.
As for the particle contents, we concentrate on 
the standard model and its minimal 
supersymmetric extension, although our discussions could proceed
in the same way for more extended models as well.

\subsection{$\dot\mu$}

First of all, we focus on the time variation of the
proton-electron mass ratio $\mu = m_{\rm p} / m_{\rm e}$. 
The time variation
of $\mu$ is given by
\beq
  \frac{\dot\mu}{\mu} = \frac{\dot{m}_{\rm p}}{m_{\rm p}} -
  \frac{\dot{m}_{\rm e}}{m_{\rm e}}. 
\eeq
Though the proton mass $m_{\rm p}$ depends not only on the QCD scale
$\Lambda_{\rm QCD}$ but also on the masses of the up quark and the down
quark, we set $m_{\rm p}$ to be proportional to $\Lambda_{\rm QCD}$
because these quark masses are much smaller than $\Lambda_{\rm QCD}$.
Then assuming $|\dot{m}_{\rm u}|, |\dot{m}_{\rm d}| \ll
|\dot{\Lambda}_{\rm QCD}|$, the time variation of $m_{\rm p}$ is given
by\footnote{Note that the light quark masses may contribute to the
proton mass, which may lead to the slight change of the coefficient in
Eq.(\ref{eq:proton}) \cite{Dent}.}
\beq
  \frac{\dot{m}_{\rm p}}{m_{\rm p}} = 
    \frac{\dot{\Lambda}_{\rm QCD}}{\Lambda_{\rm QCD}}.
\eeq

The QCD scale $\Lambda_{\rm QCD}$ can be extracted from the 
Landau pole of
the renormalization group equations as
\bea
 0= \frac{1}{\alpha_3(\Lambda_{\rm QCD})}
 &=&
   \frac{1}{\alpha_{\rm X}(M_{\rm s})}
     + \frac{b^{s}_3}{2 \pi} \ln \lmk \frac{M_{\rm s}}{M_{\rm SUSY}} \rmk
      + \frac{b_3}{2 \pi} \ln \lmk \frac{M_{\rm SUSY}}{m_{\rm t}} \rmk
          \non \\
  &&
    + \frac{b^{\rm t-b}_3}{2 \pi}
 \ln \lmk \frac{m_{\rm t}}{m_{\rm b}} \rmk
      + \frac{b^{\rm b-c}_3}{2 \pi} 
\ln \lmk \frac{m_{\rm b}}{m_{\rm c}} \rmk
      + \frac{b^{\rm c}_3}{2 \pi} 
          \ln \lmk \frac{m_{\rm c}}{\Lambda_{\rm QCD}} \rmk.
\eea
Here the parameters $b_i$ are given by $b_i\,=(b_1, b_2,b_3)=(41/10,
-19/6, -7)$, $b^{s}_i\,=(b^{s}_1, b^{s}_2, b^{s}_3)=(33/5, 1, -3)$, and
$b^{\rm t-b}_3=-23/3,\,b^{\rm b-c}_3=-25/3,\,b^{\rm
c}_3=-9$. $\alpha_{\rm X} = \alpha_{\rm X}(M_{\rm s})$ is the gauge
coupling unified at the string scale $M_{\rm s}$, $M_{\rm SUSY}$ is the
supersymmetry (SUSY) breaking scale, and $m_{\rm t}, m_{\rm b}, m_{\rm
c}$ are the masses of top, bottom, and charm quarks, respectively.
Reduction to the case of nonsupersymmetric theory would be obvious, that
is, we take $M_{\rm SUSY}=M_{\rm s}$.  Then, the time variation of the
QCD scale is given by
\bea
  \frac{b^{c\rm }_3}{2 \pi} 
\frac{\dot{\Lambda}_{\rm QCD}}{\Lambda_{\rm QCD}}
 &=&
  - \frac{\dot{\alpha}_{\rm X}}{\alpha_{\rm X}^2}
  + \frac{b^{s}_3}{2 \pi} \frac{\dot{M}_{\rm s}}{M_{\rm s}}
  + \frac{b_3 - b^{s}_3}{2 \pi} \frac{\dot{M}_{\rm SUSY}}{M_{\rm SUSY}}
  - \frac{b_3}{2 \pi} \frac{\dot{m}_{\rm t}}{m_{\rm t}}
          \non \\ 
 &&
  + \frac{b^{\rm c}_3}{2 \pi} \frac{\dot{m}_c}{m_c}
  + \frac{b^{\rm t-b}_3}{2 \pi} 
     \lmk \frac{\dot{m}_{\rm b}}{m_{\rm b}} 
          - \frac{\dot{m}_{\rm t}}{m_{\rm t}} \rmk
  + \frac{b^{\rm b-c}_3}{2 \pi} 
     \lmk \frac{\dot{m}_{\rm c}}{m_{\rm c}} 
          - \frac{\dot{m}_{\rm b}}{m_{\rm b}} \rmk.
\eea
Here and hereafter, for simplicity, we assume the universality of the
time dependence of fermion masses, that is, the time variation of all
the fundamental fermion masses except three light quarks, u, d, and
s,\footnote{The current masses of these light quarks are irrelevant to
our analysis.} is identical, which is denoted by $\dot{m}_{\rm f}/m_{\rm
f}$. Later we present a set of sufficient conditions to realize such a
universality in the context of a dilaton runaway scenario. Under such a
universality, the last two terms of the right hand side of the above
equation are dropped. Then, the time variation of the proton mass is
given by
\beq
  \frac{\dot{m}_{\rm p}}{m_{\rm p}} = 
    \frac{\dot{\Lambda}_{\rm QCD}}{\Lambda_{\rm QCD}}
  =
    \frac{2\pi}{9} \frac{\dot{\alpha}_{\rm X}}{\alpha_{\rm X}^2}
  + \frac{1}{3} \frac{\dot{M}_{\rm s}}{M_{\rm s}}
  + \frac{4}{9} \frac{\dot{M}_{\rm SUSY}}{M_{\rm SUSY}}
  + \frac{2}{9} \frac{\dot{m}_{\rm f}}{m_{\rm f}}.
\label{eq:proton}
\eeq
After all, the time variation of the proton-electron mass ratio $\mu =
m_{\rm p} / m_{\rm e}$ becomes
\beq
  \frac{\dot{\mu}}{\mu} = 
   \frac{2\pi}{9} \frac{\dot{\alpha}_{\rm X}}{\alpha_{\rm X}^2}
  + \frac{1}{3} \frac{\dot{M}_{\rm s}}{M_{\rm s}}
  + \frac{4}{9} \frac{\dot{M}_{\rm SUSY}}{M_{\rm SUSY}}
  - \frac{7}{9} \frac{\dot{m}_{\rm f}}{m_{\rm f}},
 \label{eq:mu}
\eeq
where we have used the universality of the time variation of fermion masses
$\dot{m}_{\rm e}/m_{\rm e}=\dot{m}_{\rm f}/m_{\rm f}$.

\subsection{$\dot\alpha$}

Next, we discuss the time variation of the fine structure constant
 $\alpha$. The
renormalization group equations yield the scale dependence of the gauge
couplings $\alpha_i$ ($i$ = 1,2)
\beq
  \frac{1}{\alpha_i(M_{\rm EW})} =
   \frac{1}{\alpha_{\rm X}}
      + \frac{b^{s}_i}{2 \pi} \ln \lmk \frac{M_{\rm s}}{M_{\rm SUSY}} \rmk
      + \frac{b_i}{2 \pi} \ln \lmk \frac{M_{\rm SUSY}}{M_{\rm EW}} \rmk,
  \label{eq:alpha}
\eeq
where $M_{\rm EW} \simeq M_{\rm Z}$ is the electroweak symmetry breaking
scale. Assuming $SU(5)$ GUT, for example,  the fine structure constant
$\alpha$ can be related to the gauge couplings $\alpha_i$ ($i$ = 1,2) at
any scale as
\bea
  \frac{1}{\alpha(M_{\rm EW})} &=& 
    \frac{5}{3\alpha_1(M_{\rm EW})} + \frac{1}{\alpha_2(M_{\rm EW})} 
     \non \\
  &=&
   \frac83 \frac{1}{\alpha_{\rm X}}
      + \frac{6}{\pi} \ln \lmk \frac{M_{\rm s}}{M_{\rm SUSY}} \rmk
      + \frac{11}{6 \pi} \ln \lmk \frac{M_{\rm SUSY}}{M_{\rm EW}} \rmk ,
\eea
for $M_{\rm EW} < M_{\rm SUSY}$. After the electroweak symmetry
breaking, charged fields acquire their masses. Therefore, taking their
mass thresholds into account, the fine structure constant $\alpha$ is given
by
\beq
  \frac{1}{\alpha}  =  
   \frac{1}{\alpha (M_{\rm EW})}
    + \frac{b_{\alpha}}{2 \pi} 
         \ln \lmk \frac{M_{\rm EW}}{m_{\rm t}} \rmk
    + \sum_{f_i} \frac{b^{{\rm f}_{i}-{\rm f}_{i+1}}_{\alpha}}{2 \pi} 
         \ln \lmk \frac{m_{{\rm f}_{i}}}{m_{{\rm f}_{i+1}}} \rmk .
\eeq
%
Here, the third term in the right hand side corresponds to 
fermion mass thresholds, where $\rm f_i= t,b,\cdots,e$, 
$b_{\alpha} = 32/3$ 
and $b_\alpha^{{\rm f}_{i} - {\rm f}_{i+1}}$ denotes beta-function 
coefficients for fermion mass thresholds between ${\rm f}_i$ and 
${\rm f}_{i+1}$.

Then, the time variation of the fine structure constant is given by
\beq
  \frac{\dot{\alpha}}{\alpha^2}
  = \frac{80}{27} \frac{\dot{\alpha}_{\rm X}}{\alpha_{\rm X}^2}
  - \frac{50}{9\pi} \frac{\dot{M}_{\rm s}}{M_{\rm s}}
  + \frac{257}{54\pi} \frac{\dot{M}_{\rm SUSY}}{M_{\rm SUSY}}
  - \frac{7}{2\pi} \frac{\dot{M}_{\rm EW}}{M_{\rm EW}}
  + \frac{116}{27\pi} \frac{\dot{m}_{\rm f}}{m_{\rm f}},
\label{time-v-alpha}
\eeq
where we have used the universality of the time variation of fermion
masses except three light quarks u, d, s, and $m_{\rm u} \simeq m_{\rm
d} \simeq m_{\rm s} \simeq \Lambda_{\rm QCD}$ because light quarks are
confined and effectively have a mass of the order of $\Lambda_{\rm
QCD}$.

We have assumed that the gauge couplings are unified to $\alpha_X$.
The gauge coupling unification is consistent with experimental 
values on the gauge couplings within the framework of 
the minimal supersymmetric standard model, but not in 
the (non-SUSY) standard model.
For the latter case, we need some corrections to gauge couplings 
at $M_{\rm s}$.
Such corrections can appear from gauge kinetic functions, which 
depend on moduli fields other than the dilaton.
We assume that moduli-dependent corrections to the gauge couplings 
do not vary while only the dilaton varies in the time range 
relevant to our analysis.

\subsection{Universality of Time Variation of Fermion Masses}

The four dimensional effective low-energy action related to the
generation of  fermion masses is given by
\bea
  S &\supset& \int d^4 x \sqrt{\widetilde{g}} 
    \lkk 
        B_{H_{\rm u}}(\phi)\,
         (D_{\mu} \widetilde{H_{\rm u}})^{\dagger} 
          D^{\mu} \widetilde{H_{\rm u}}
      + B_{H_{\rm d}}(\phi)\,
         (D_{\mu} \widetilde{H_{\rm d}})^{\dagger} 
          D^{\mu} \widetilde{H_{\rm d}} \right. \non \\
    && \qquad \qquad \quad
      + i B_{\rm Q_{\rm j}} \overline{\widetilde{Q_{\rm j}}} 
            \slaD \widetilde{Q_{\rm j}} 
      + i B_{\rm u_{\rm j}} \overline{\widetilde{u_{\rm j}}} 
            \slaD \widetilde{u_{\rm j}} 
      + i B_{\rm d_{\rm j}} \overline{\widetilde{d_{\rm j}}} 
            \slaD \widetilde{d_{\rm j}} 
      + i B_{\rm L_{\rm j}} \overline{\widetilde{L_{\rm j}}} 
            \slaD \widetilde{L_{\rm j}} 
      + i B_{\rm e_{\rm j}} \overline{\widetilde{e_{\rm j}}} 
            \slaD \widetilde{e_{\rm j}} \non \\
    && \qquad \qquad \quad \left.
      + i B_{y_{\rm u_{\rm j}}} \widetilde{y_{\rm u_{\rm j}}}  
         \overline{\widetilde{Q_{\rm j}}} \widetilde{u_{\rm j}} 
         \widetilde{H_{\rm u}}
      + i B_{y_{\rm d_{\rm j}}} \widetilde{y_{\rm d_{\rm j}}}  
         \overline{\widetilde{Q_{\rm j}}} \widetilde{d_{\rm j}} 
         \widetilde{H_{\rm d}}
      + i B_{y_{\rm e_{\rm j}}} \widetilde{y_{\rm e_{\rm j}}}  
         \overline{\widetilde{L_{\rm j}}} \widetilde{e_{\rm j}} 
         \widetilde{H_{\rm d}}
    \rkk,
\eea
where $\slaD = \gamma^{\mu} D_{\mu}$, $D_{\mu}$ represents the covariant
derivative, and ${\rm j}$ is the generation index. Canonically
normalizing all fields yield effective Yukawa couplings and fermion
masses,
\bea
  y_{\rm u_{\rm j}} &=& \widetilde{y_{\rm u_{\rm j}}} 
   \frac{B_{y_{\rm u_{\rm j}}}}
   {\sqrt{ B_{\rm Q_{\rm j}}B_{\rm u_{\rm j}}B_{H_{\rm u}} }}, 
  \qquad\qquad
  m_{\rm u_{\rm j}} = \frac{y_{\rm u_{\rm j}} v_{\rm u}}{\sqrt{2}},
      \non \\
  y_{\rm d_{\rm j}} &=& \widetilde{y_{\rm d_{\rm j}}} 
   \frac{B_{y_{\rm d_{\rm j}}}}
   {\sqrt{ B_{\rm Q_{\rm j}}B_{\rm d_{\rm j}}B_{H_{\rm d}} }},
  \qquad\qquad
  m_{\rm d_{\rm j}} = \frac{y_{\rm d_{\rm j}} v_{\rm d}}{\sqrt{2}},
      \non \\
  y_{\rm e_{\rm j}} &=& \widetilde{y_{\rm e_{\rm j}}} 
   \frac{B_{y_{\rm e_{\rm j}}}}
   {\sqrt{ B_{\rm L_{\rm j}}B_{\rm e_{\rm j}}B_{H_{\rm d}} }},
  \qquad\qquad
  m_{\rm e_{\rm j}} = \frac{y_{\rm e_{\rm j}} v_{\rm d}}{\sqrt{2}}.
\eea
%
Uplike and downlike fermions acquire masses $m_{\rm u_{\rm j}} = y_{\rm
u_{\rm j}} v_{\rm u} / \sqrt{2}$ and $m_{\rm d_{\rm j}} = y_{\rm d_{\rm
j}} v_{\rm d} / \sqrt{2}$ through the Higgs mechanism, respectively.
Here $y_{\rm i}$ is a Yukawa coupling constant and $v_{\rm u}$ and
$v_{\rm d}$ are the VEV of the up-type and down-type Higgs fields,
respectively.  Note that $v_{\rm u}(v_{\rm d})$ is replaced by the VEVs
of the standard Higgs field $v$ in the case of the nonsupersymmetric
standard model with a single Higgs doublet.

Then, the time variation of fermion masses is given by
\bea
  \frac{\dot{m}_{\rm u_{\rm j}}}{m_{\rm u_{\rm j}}} &=&  
    \frac{\dot{y}_{\rm u_{\rm j}}}{y_{\rm u_{\rm j}}} 
   + \frac{\dot{v}_{\rm u}}{v_{\rm u}}, \non \\
  \frac{\dot{m}_{\rm d_{\rm j}}}{m_{\rm d_{\rm j}}} &=&  
    \frac{\dot{y}_{\rm d_{\rm j}}}{y_{\rm d_{\rm j}}} 
   + \frac{\dot{v}_{\rm d}}{v_{\rm d}}, \non \\
  \frac{\dot{m}_{\rm e_{\rm j}}}{m_{\rm e_{\rm j}}} &=&  
    \frac{\dot{y}_{\rm e_{\rm j}}}{y_{\rm e_{\rm j}}} 
   + \frac{\dot{v}_{\rm d}}{v_{\rm d}}.
\eea 
Thus, the universality of the time variation of fermion masses is 
realized, for example, 
if the following conditions are satisfied,
\bea
  \frac{\dot{y}_{\rm u_{\rm j}}}{y_{\rm u_{\rm j}}} 
   = \frac{\dot{y}_{\rm d_{\rm k}}}{y_{\rm d_{\rm k}}} 
   = \frac{\dot{y}_{\rm e_{\rm l}}}{y_{\rm e_{\rm l}}} 
   \equiv \frac{\dot{y}_{\rm f}}{y_{\rm f}},
  \qquad
  \frac{\dot{v}_{\rm u}}{v_{\rm u}}
   = \frac{\dot{v}_{\rm d}}{v_{\rm d}}
   = \frac{\dot{v}}{v},
  \label{eq:ycond}
\eea
where ${\rm j}, {\rm k}, {\rm l}$ are generation indices and the last
equality comes from $v^2 = v_{\rm u}^2+ v_{\rm d}^2$. Note that the
second condition is unnecessary in the case of the nonsupersymmetric
minimal standard model.  Here, we have implicitly assumed that the scale
dependence of the Yukawa couplings is negligible, that is, the time
dependence of the Yukawa couplings is dominated by their dilaton
dependence.

Here, we comment on radiative corrections on Yukawa couplings.
Renormalization group effects on Yukawa couplings are the same 
among quarks except the top quark.
Thus, radiative corrections do not violate the universal 
time variation of Yukawa couplings among quarks except the 
top quark.
Furthermore, the mass ratios of other quarks to the top quark 
do not change drastically between $M_{\rm s}$ and $M_{\rm Z}$, {\it i.e.}
$m_{\rm f}(M_{\rm s})/m_{\rm t}(M_{\rm s}) \sim m_{\rm f}(M_{\rm
  Z})/m_{\rm t}(M_{\rm Z})$.
Hence, even including radiative corrections,
the universal time variation of quark Yukawa couplings 
would be a reasonable assumption, 
and such corrections on the time variation on $\Lambda_{\rm QCD}$
would be sufficiently small.
The same discussion holds true for 
the universal time variation of Yukawa couplings 
only among leptons.
However, radiative corrections on quark Yukawa couplings 
are different from those on lepton Yukawa couplings, because 
of corrections from $\alpha_3$.
Such difference would be estimated as 
$\dot Y_q/Y_q - \dot Y_\ell / Y_\ell = a \dot \alpha_X$, 
where $|a| \leq O(1)$, 
and it has some effect on $\dot \alpha/\alpha$ Eq.(\ref{time-v-alpha}), 
but it can be neglected compared with the first term in 
Eq.~(\ref{time-v-alpha}).

\subsection{Electroweak scale, Higgs VEV, and SUSY scale}

Before investigating the time dependence of the VEV of the Higgs fields
$\dot{v}/v$, we discuss the time dependence of the electroweak symmetry
breaking scale $M_{\rm EW}$ characterized by the gauge boson
mass $M_{\rm Z}$,
\beq
  M_{\rm EW} \simeq M_{\rm Z} 
     = \frac{v}{2} \sqrt{g^2+g'^2}
     = v \sqrt{\pi \lmk \frac35 \alpha_1 +\alpha_2 \rmk},
  \label{eq:ewscale}
\eeq      
where $\alpha_i = \alpha_i(M_{\rm EW})$, and $g$ and $g'$ are
$SU(2)_{\rm L}$ and $U(1)_{\rm Y}$ gauge coupling constants
respectively.
Then, the time variation of the electroweak symmetry breaking scale
$\dot{M}_{\rm EW}/{M_{\rm EW}}$ is given by
\beq
  \frac{{\dot M}_{\rm EW}}{M_{\rm EW}} 
    = \frac{3\dot{\alpha_1}+5\dot{\alpha_2}}{2(3\alpha_1+5\alpha_2)}
      + \frac{\dot{v}}{v}.
  \label{eq:ew}
\eeq 
Inserting Eq. (\ref{eq:alpha}) into this equation  yields
\beq
  \frac{\dot{M}_{\rm EW}}{M_{\rm EW}} 
\simeq  \frac{3\alpha_1^2+5\alpha_2^2}{2(3\alpha_1+5\alpha_2)} 
           \frac{\dot{\alpha}_{\rm X}}{\alpha_{\rm X}^2}
    - \frac{99\alpha_1^2+25\alpha_2^2}{20\pi(3\alpha_1+5\alpha_2)} 
           \frac{\dot{M}_s}{M_s}
    + \frac{45\alpha_1^2+125\alpha_2^2}{24\pi(3\alpha_1+5\alpha_2)} 
           \frac{\dot{M}_{\rm SUSY}}{M_{\rm SUSY}}
    + \frac{\dot{v}}{v}
\simeq \frac{\dot{v}}{v}, 
\eeq
where we have used 
$\alpha_1 \sim 1/60 \ll 1$ and $ \alpha_2 \sim 1/30 \ll
1$.

We now investigate the time variation of the VEV of the Higgs fields,
$\dot{v}/v$. First, we consider the nonsupersymmetric minimal standard
model. The Lagrangian density related to the standard Higgs field is
expected to read
\bea
  \CL &\supset& 
         B_{\rm H}(\phi)\,
         (D_{\mu} \widetilde{H})^{\dagger} 
          D^{\mu} \widetilde{H}
       - B_{\lambda}(\phi)\,\widetilde{\lambda}
         \lmk \widetilde{H}^{\dagger} \widetilde{H}
              - \frac{\widetilde{v}^2}{2}  
         \rmk^2 \non \\      
      &=&  
         (D_{\mu} H)^{\dagger} D^{\mu} H
       - \lambda
           \lmk H^{\dagger} H - \frac{v^2}{2}  
         \rmk^2,
\eea
with $H \equiv \sqrt{B_{\rm H}} \widetilde{H}, \lambda \equiv B_{\lambda}
\widetilde{\lambda} / B_{\rm H}^2,$ and $ v^2 \equiv B_{\rm H}
\widetilde{v}^2$. Assuming that $\widetilde{v}$ is intrinsic and has no
time dependence, the time variation of the VEV of the standard Higgs field
is given by
\beq
  \frac{\dot{v}}{v} = \frac12 \frac{\dot{B_{\rm H}}}{B_{\rm H}}.
  \label{eq:v}
\eeq

Next, we consider the SUSY model.
The neutral components of up and down sector Higgs fields, 
$h_u$ and $h_d$, have the following potential,
\begin{equation}
V =  m_1^2|h_d|^2 + m_2^2|h_u|^2 + b (h_dh_u + h.c.) 
+ \frac18 (g^2+g'^2)\left( |h_d|^2 - |h_u|^2 \right)^2,
\end{equation}
with 
\begin{equation}
m_1^2 = m_{H_d}^2 + \mu_H^2, \qquad 
m_2^2 = m_{H_u}^2 + \mu_H^2,
\end{equation}
where $m_{H_{u,d}}^2$ are SUSY breaking scalar masses squared of 
Higgs fields, $h_{u,d}$, and $\mu_H$ is the supersymmetric mass 
parameter.
In addition, the parameter $b$ is also SUSY breaking 
parameter with mass dimension two, that is, the so-called 
$b$-term.

By using the stationary conditions, 
$\partial V/\partial h_{u,d}=0$, we obtain 
\begin{eqnarray}
 & & 
\frac14(g^2+g'^2)v^2 = -m_1^2-m_2^2 
+\frac{\tan^2 \beta +1}{\tan^2 \beta -1} (m_1^2 -m_2^2), \\
 & & -b (\tan \beta + \cot \beta) = m_1^2+m_2^2 ,
\end{eqnarray}
where $\tan \beta = v_{\rm u}/v_{\rm d}$.
Furthermore, for a moderate and/or large value of $\tan \beta$, 
{\it i.e.} 
$\tan^2 \beta >O(1)$, these equations reduce to 
\begin{eqnarray}
 & & 
\frac18(g^2+g'^2)v^2 = -m_2^2 , \\
 & & -b \tan \beta  = m_1^2+m_2^2. 
\end{eqnarray}
That implies that if the time variation of mass parameters is 
the same, i.e.
\begin{equation}
\frac{d}{dt} \ln \mu^2_H= \frac{d}{dt} \ln m_{H_u}^2=
\frac{d}{dt} \ln m_{H_d}^2= \frac{d}{dt} \ln b ,
\label{mass-t-vary}
\end{equation}
$\tan \beta$ does not vary, that is, 
$\dot v_{\rm u}/v_{\rm u} = \dot v_{\rm d}/v_{\rm d}$.
The above assumption (\ref{mass-t-vary}) might be plausible 
for mass parameters at tree level, but $m_{H_u}^2$ has 
a significant radiative correction due to stop mass, 
\begin{equation}
\delta m_{Hu}^2 \sim - \frac{3y_t^2m_{\tilde t}^2}{4 \pi^2}
\ln(M_s /m_{\tilde t}),
\end{equation}
where $m_{\tilde t}$ is SUSY breaking stop mass.
Thus, in general, the value of $\tan \beta$ varies in time, and 
the time variations of $\dot v_{\rm u}/v_{\rm u}$ and 
$\dot v_{\rm d}/v_{\rm d}$ are different.
To take into account this aspect, we have to consider the 
situation that the time dependence of up-type quark masses 
are different from those of down-type quark masses and lepton
masses, and we have to introduce another parameter to represent 
such difference.
Such extension of our analysis is straightforward and 
 would enlarge a favorable parameter space.
(Note that because of $\dot M_{\rm SUSY}/M_{\rm SUSY}$
the SUSY model has more degrees of freedom than the non-SUSY 
standard model.)
Similarly, the time variation of $v$ also depends on 
those of several values, $\mu_H$, $m^2_{H_u}$, $y_t$, $m_{\tilde t}^2$
as well as the gauge couplings.
To simplify our analysis, we use the same parameterization 
as the non-SUSY model Eq.(\ref{eq:v}).

Now let us discuss the time dependence of the SUSY breaking scale 
$M_{\rm SUSY}$.
Although it 
strongly depends on the SUSY breaking model, we give one example based
on the gaugino condensation and gravity mediation model. We consider a
hidden sector, in which a hidden gauge group with a coupling
$\alpha_{\rm h}$ blows up and hence gauginos $\lambda^a$ condensate at
some scale $M_{\rm c}$, which breaks the SUSY. Then, 
repeating the same argument as the case of the QCD scale,
 the condensation scale $M_{\rm c}$ is given
through the RG flow by
\beq
  M_{\rm c} = M_{\rm s} e^{-\frac{2\pi}{\alpha_{\rm h}(M_{\rm s}) b_{\rm h}}},
  \qquad \la \lambda^a \lambda^a \ra = M_{\rm c}^3 \ne 0,
\eeq
where $b_{\rm h}$ is the beta-function coefficient which, 
for example, is given by
$b_{\rm h} = 3 N_c$ for the gauge group $SU(N_c)$. If this breaking is
transmitted to the visible sector through the gravitational interaction,
the SUSY breaking scale $M_{\rm SUSY}$ is given by
\beq
  M_{\rm SUSY} = 8\pi G M_{\rm c}^3 = \frac{M_{\rm c}^3}{M_{G}^2}. 
\eeq
In this case, the time variation of the SUSY breaking scale is given by
\beq
  \frac{\dot{M}_{\rm SUSY}}{M_{\rm SUSY}}
    = 3 \frac{\dot{M}_{\rm c}}{M_{\rm c}}
     - 2 \frac{\dot{M}_{G}}{M_{G}} 
    = - \frac{6\pi}{b_{\rm h}} 
    \frac{\dot{\alpha}_{\rm h}(M_{\rm s})}{\alpha_{\rm h}^2(M_{\rm s})}
     - 2 \frac{\dot{M}_{G}}{M_{G}} .
  \label{eq:SUSY}
\eeq
%

\section{Time variation in a runaway dilaton scenario}

\subsection{Runaway Dilaton}

Now, we estimate the time variation of the 
proton-electron mass ratio and the fine
structure constant in the context of a runaway dilaton scenario. From
the four dimensional effective low-energy action (\ref{eq:action}), we
have the following relations,
\bea
  && B_{g}(\phi) M_{\rm s}^2 = \frac{1}{16\pi G}, \non \\
  && B_{F}(\phi) = \frac{1}{8\pi \alpha_{\rm X}}.
\eea
Since the coupling functions are given in Eq. (\ref{eq:cf}), the dilaton
dependence of the gravitational coupling $G$ and the unified gauge
coupling $\alpha_{\rm X}$ is given by
\bea
  && M_G^{2} = (8\pi G)^{-1} = 2 M_{\rm s}^2 B_g(\phi) 
     = 2 M_{\rm s}^2 C_g (1 + d_g e^{-\phi}), \non \\  
  && \alpha_{\rm X}^{-1} = 8 \pi B_F(\phi) 
     = 8 \pi C_F (1 + d_F e^{-\phi}),
\eea
which leads to
\bea
  \frac{\dot{M_G}}{M_G} &=& 
      - \frac{d_g\,e^{-\phi}}{2(1+d_g e^{-\phi})} \dot{\phi}
      + \frac{\dot{M}_{\rm s}}{M_{\rm s}} 
                        \simeq 
      - \frac12\,d_g\,e^{-\phi} \dot{\phi}
      + \frac{\dot{M}_{\rm s}}{M_{\rm s}}, \non \\
  \frac{\dot{\alpha}_{\rm X}}{\alpha_{\rm X}^2}
    &=& 8 \pi A_F e^{-\phi} \dot{\phi} > 0,
  \label{eq:alphax}
\eea
where we set $\dot{\phi} > 0$ without generality and we have assumed
$e^{-\phi} \ll 1$.  We regard
the string scale $M_{\rm s}$ as fundamental and hence it has no time
dependence $\dot{M_{\rm s}}=0$.

In the context of the dilaton
runaway scenario, the sufficient
 conditions  for the universality of the time 
variation of Yukawa couplings, (\ref{eq:ycond}), are satisfied,
for example, in the case that the dilaton dependent functions have the
following properties,
\bea
  && B_{\rm Q_{\rm j}} = B_{\rm L_{\rm k}} \equiv B_{\rm D}
     = C_{\rm D} + A_{\rm D} e^{-\phi} 
     = C_{\rm D} (1 + d_{\rm D} e^{-\phi}), \non \\
  && B_{\rm u_{\rm j}} = B_{\rm d_{\rm k}} = B_{\rm e_{\rm l}}
       \equiv B_{\rm S}
     = C_{\rm S} + A_{\rm S} e^{-\phi} 
     = C_{\rm S} (1 + d_{\rm S} e^{-\phi}), \non \\
  && B_{\rm y_{\rm u_{\rm j}}} = B_{\rm y_{\rm d_{\rm k}}} 
       = B_{\rm y_{\rm e_{\rm l}}} \equiv B_{\rm y}
     = C_{\rm y} + A_{\rm y} e^{-\phi} 
     = C_{\rm y} (1 + d_{\rm y} e^{-\phi}), \non \\
  && B_{\rm H_{\rm u}} = B_{\rm H_{\rm d}} \equiv B_{\rm H}
     = C_{\rm H} + A_{\rm H} e^{-\phi} 
     = C_{\rm H} (1 + d_{\rm H} e^{-\phi}).
  \label{eq:bcond}
\eea 
Hereafter we assume that the dilaton dependent functions satisfy the
above conditions (\ref{eq:bcond}). In this case, the universal time
variation of fermion masses $\dot{m}_{\rm f}/m_{\rm f}$ is given by
\beq
  \frac{\dot{m}_{\rm f}}{m_{\rm f}} = 
     \frac{\dot{y}_{\rm f}}{y_{\rm f}} + \frac{\dot{v}}{v},
\label{eq:mf2}
\eeq
where the universal time variation of
Yukawa couplings $y_{\rm f}$ is estimated as
\beq
  \frac{\dot{y}_{\rm f}}{y_{\rm f}} \simeq 
    - d\,e^{-\phi} \dot{\phi} ,
  \label{eq:yukawa}
\eeq
with 
\beq
d \equiv d_{\rm y} - (d_{\rm D} + d_{\rm S} + d_{\rm H})/2,
\label{eq:total-d}
\eeq
and (\ref{eq:v}) reads
\beq
  \frac{\dot{v}}{v} 
          = - \frac{d_{\rm H}\,e^{-\phi}}{2(1+d_{\rm H} e^{-\phi})} 
            \dot{\phi}
          \simeq - \frac12\,d_{\rm H}\,e^{-\phi} 
            \dot{\phi}.
  \label{eq:vdilaton}
\eeq
Note that $d$ can take either a positive value or a negative one, 
which is an important point to account for the 
decline of $\mu$. 

After all, 
the universal time variation of
fermion masses is given by
\beq
  \frac{\dot{m}_{\rm f}}{m_{\rm f}} 
    = \frac{\dot{y}_{\rm f}}{y_{\rm f}} + \frac{\dot{v}}{v}
    \simeq - \lmk d+\frac{d_{\rm H}}{2} \rmk\,e^{-\phi} 
            \dot{\phi}.
  \label{eq:mf}
\eeq

\subsection{$\dot\mu$ and $\dot\alpha$ in the Runaway Dilaton Scenario}

Finally, we show that the choice of adjustable parameters allows them to
fit the observed time variation of the proton-electron mass ratio and
the fine structure constant in the dilaton runaway scenario, and give
constraints on the parameters.

In the case of the  non-SUSY model (${\dot M_{\rm SUSY}}=0$), the time
variation of the proton-electron mass ratio 
$\mu = m_{\rm p} / m_{\rm e}$ reads
\bea
  \frac{\dot{\mu}}{\mu} 
   &=& 
    \frac{2\pi}{9} \frac{\dot{\alpha}_{\rm X}}{\alpha_{\rm X}^2}
    - \frac{7}{9} \frac{\dot{m}_{\rm f}}{m_{\rm f}} \non \\
   &\simeq&
    \frac{2\pi}{9}\,e^{-\phi} \dot{\phi}\,
      \lkk 
        8\pi A_F + \frac{7}{2\pi} \lmk d + \frac{d_{\rm H}}{2} \rmk
      \rkk    ,
  \label{eq:nonmu}
\eea
where we have used Eqs.\ (\ref{eq:alphax}) and (\ref{eq:mf}) in the
second equality. As is seen in  Eq.\ (\ref{eq:total-d}), 
$d$ can be negative. 
On the other hand, the time variation of the fine structure
constant is given by
\bea
  \frac{\dot{\alpha}}{\alpha^2}
  &=& 
       \frac{80}{27} \frac{\dot{\alpha}_{\rm X}}{\alpha_{\rm X}^2}
     - \frac{7}{2\pi} \frac{\dot{M}_{\rm EW}}{M_{\rm EW}}
     + \frac{116}{27\pi} \frac{\dot{m}_{\rm f}}{m_{\rm f}}, \non \\
  &\simeq&
       \frac{80}{27} \frac{\dot{\alpha}_{\rm X}}{\alpha_{\rm X}^2}
     - \frac{7}{2\pi} \frac{\dot{v}}{v}
     + \frac{116}{27\pi} \frac{\dot{m}_{\rm f}}{m_{\rm f}}, \non \\
  &\simeq&
    \frac{80}{27}\,e^{-\phi} \dot{\phi}\,
      \lkk 
    8\pi A_F - \frac{29}{20\pi} \lmk d + \frac{43}{464} d_{\rm H} \rmk
      \rkk.
  \label{eq:nonalpha}
\eea
As mentioned in the introduction,
observations indicate 
$\dot{\alpha}/\alpha >
0$ and $\dot{\mu}/\mu < 0$, 
which impose constraints on the parameters,
\beq
  \frac{29}{160\pi^2} \lmk d + \frac{43}{464} d_{\rm H} \rmk
    \lesssim A_{F}
    \lesssim
  - \frac{7}{16\pi^2} \lmk d + \frac{d_{\rm H}}{2} \rmk.
\eeq
These constraints can be easily satisfied if $d < 0,~\abs{d} \gg d_{\rm
H}$ and $A_F \lesssim 29\abs{d}/(160\pi^2)$. Though the time variations
of the proton-electron mass ratio and the fine structure constant depend
on the evolution of the dilaton field $\phi$, from Eqs.\
(\ref{eq:dalpha}), (\ref{eq:dmu}), (\ref{eq:nonmu}), and
(\ref{eq:nonalpha}), we can expect that observed variation can be fitted
by taking our model parameters appropriately around order of unity.

In the case of the SUSY model, the time variation of the proton-electron
mass ratio $\mu = m_{\rm p} / m_{\rm e}$ becomes
\bea
 \frac{\dot{\mu}}{\mu}
  &=&
   \frac{2\pi}{9} \frac{\dot{\alpha}_{\rm X}}{\alpha_{\rm X}^2}
   + \frac{4}{9} \frac{\dot{M}_{\rm SUSY}}{M_{\rm SUSY}}
   - \frac{7}{9} \frac{\dot{m}_{\rm f}}{m_{\rm f}} \non \\
  &\simeq&
   \frac{2\pi}{9}\,e^{-\phi} \dot{\phi}\,
     \lkk
         8\pi A_F \lmk 1 -12 \frac{A_{\rm h}}{A_F b_{\rm h}} \rmk
       + \frac{7}{2\pi} \lmk d+ \frac{d_H}{2} +\frac{4d_{g}}{7} \rmk
     \rkk,
 \label{eq:SUSYmu}
\eea
where we have used Eqs.\ (\ref{eq:SUSY}),
 (\ref{eq:alphax}) and (\ref{eq:mf}). On the other hand, time variation
of the fine structure constant is given by
\bea
 \frac{\dot{\alpha}}{\alpha^2}
 &=&
      \frac{80}{27} \frac{\dot{\alpha}_{\rm X}}{\alpha_{\rm X}^2}
    + \frac{257}{54\pi} \frac{\dot{M}_{\rm SUSY}}{M_{\rm SUSY}}
    - \frac{7}{2 \pi} \frac{\dot{M}_{\rm EW}}{M_{\rm EW}}
    + \frac{116}{27\pi} \frac{\dot{m}_{\rm f}}{m_{\rm f}} \non \\
 &\simeq&
   \frac{80}{27}\,e^{-\phi} \dot{\phi}\,
     \lkk
         8\pi A_F \lmk 1
         - \frac{771}{80} \frac{A_{\rm h}}{A_F b_{\rm h}} \rmk
       - \frac{29}{20\pi} \lmk d + \frac{43}{464}d_H 
           -\frac{257}{232} d_{g} \rmk
     \rkk.
 \label{eq:SUSYalpha}
\eea
In this case, too, from Eqs. (\ref{eq:dalpha}), (\ref{eq:dmu}),
(\ref{eq:SUSYmu}), and (\ref{eq:SUSYalpha}), we can expect that our
parametrization can fit the observed variation naturally. In fact, the
SUSY model has more degrees of freedom than the non-SUSY standard model
because of $\dot M_{\rm SUSY}/M_{\rm SUSY}$. Furthermore, if we
introduce the time variation of the ratio of $\dot v_u/v_u$ to $\dot
v_d/v_d$ as well, we have a wider parameter space to fit the observed
variation.

\section{Comparison with observations}

Having formulated the time variation of the fundamental constants
in the particle-physics context and given their explicit form in
the dilaton runaway scenario, we now solve cosmological evolution
of the dilaton field to show our model can account for the 
observed variation.  Before giving an explicit result, however,
we must consider other experimental consequences of the dilaton
coupling which impose a stringent constraints on the parameter space.

\subsection{Experimental Constraints on Dilaton Coupling}

The dilaton coupling to hadronic matter induces deviations 
from general relativity: post-Newtonian deviations
 from general relativity 
and the violations of the equivalence principle \cite{DP,DPV}. 
After integration by parts, the action of the dilaton is rewritten as
\beq
S= \int d^4 x \sqrt{\widetilde{g}} 
    \lmk \frac{B_g (\phi)}{\alpha'} \, \widetilde{R} 
           - \frac{ Z(\phi)}{\alpha'} \,
           (\widetilde{\nabla} \phi)^2 
            - V(\phi) 
           + \cdots 
    \rmk\, ,
\label{eq:action-dilaton}
\eeq
where $Z(\phi)=C_{\phi}-A_{\phi}e^{-\phi}$ and 
$B_g(\phi)=C_g(1+d_ge^{-\phi})$. 

From the effective mass $m(\phi)$ of a test particle (composed of
hadronic matter) in the Einstein frame metric
$g_{\mu\nu}=B_g(\phi)\widetilde{g}_{\mu\nu}$, using
Eq.(\ref{eq:proton}), the strength of the coupling of the dilaton to
hadronic matter, $\alpha_{had}$, is given by \cite{DP,DPV} \bea
\alpha_{had}\simeq \sqrt{\frac{2B_g(\phi)}{Z(\phi)}}\frac{d\ln
m(\phi)}{d\phi} &\simeq & \sqrt{\frac{2B_g(\phi)}{Z(\phi)}}\, \lkk
\frac{d\ln m_{\rm p}(\phi)}{d\phi}-\frac{1}{2}\frac{d\ln
B_g(\phi)}{d\phi} \rkk \nonumber\\ &\simeq
&e^{-\phi}\sqrt{\frac{2C_g}{C_{\phi}}}\, \lkk \frac{16\pi^2}{9} A_F -
\frac{2}{9} \lmk d + \frac{d_{\rm H}}{2} \rmk +\frac{1}{2}d_g \rkk ,
\label{alphahad} \eea where the non-SUSY standard model is assumed in
the second line.  Since the mass of a test particle depends on the
dilaton, the test particle experiences an acceleration, $-{\bf \nabla}
\ln m$, in addition to the usual free-fall acceleration ${\bf g}$, which
results in the violation of the universality of free-fall.
$\alpha_{had}$ is related to the Eddington parameter $\gamma$ measuring
a post-Newtonian deviation from general relativity and to the E\"otv\"os
ratio $\eta$ measuring the difference in accelerations, $a_I$, between
the two test masses ($I=A,B$) \cite{DP,DPV}:
\bea
&&\gamma -1=-\frac{2\alpha_{had}^2}{1+\alpha_{had}^2}\simeq
-2\alpha_{had}^2 ,\\
&&\eta\equiv 2\frac{a_A-a_B}{a_A+a_B}\simeq 5.2\times 10^{-5}\alpha_{had}^2
\simeq -2.6\times 10^{-5}(\gamma -1). \label{eta}
\eea
The present experimental constraints are $\gamma-1=(2.1\pm 2.3)\times
10^{-5}$ \cite{cassini} and, $\eta=(-1.9\pm 2.5)\times 10^{-12}$ for
$A=$Be and $B=$Cu \cite{su}, and $\eta=(0.1\pm 4.4) \times
10^{-13}$ from the measurement of the composite-dependent differential
acceleration of Earth and Moon toward the Sun, which are made of several
materials such as Fe, Ni, Si, Mg, O\cite{Baessler}.


\begin{figure}

\includegraphics[width=13cm]{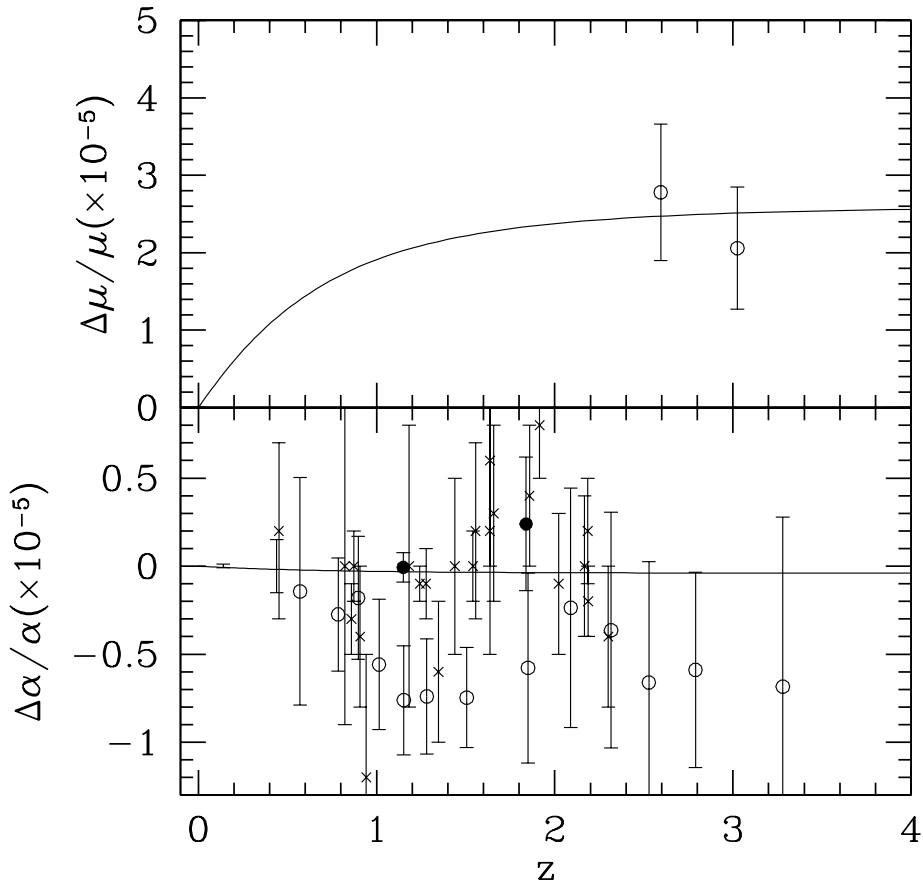} 
\caption{The time evolution of $\mu$ and $\alpha$ (solid curve) in
runaway dilaton scenario.  The upper panel is $\mu$ determined from the
spectral lines of hydrogen molecules.  The data are taken from
\cite{mpme}.  The lower panel is $\alpha$ determined from quasar
absorption lines at several redshifts. Open circles are the binned data from \cite{webb} 
(in each bin the value of $\Delta\alpha/\alpha$ is the weighted mean), 
crosses are from \cite{SCP} and filled circles are from \cite{LCM}.
The datum at $z\simeq 0.16,0.44$ is the Oklo bound \cite{DD} and the $^{187}{\rm Re}$ 
bound \cite{fujii}, respectively.}
\label{fig1}
\end{figure}

\subsection{Dilaton Evolution}

We numerically solve the cosmological evolution of the dilaton to give a
concrete example which can account for the time variations of $\mu$ and
$\alpha$. Its dynamics is determined by specifying the dilaton potential
$V(\phi)$.  We assume the following form of $V(\phi)$ in accord with
Eq.\ (\ref{eq:cf}): $V(\phi)=V_0(1+d_ve^{-\phi})$.  We further assume
that the magnitude of $V_0$ is similar to that of the cosmological
constant, $V_0\simeq 10^{-120}M_G^4$, which is miniscule in the unit of
the string scale in order for the dilaton to play the role of dark
energy.\footnote{We expect that the value of $V_0$ may be determined by
the expectation value of other fields $\chi$ which are orthogonal to the
dilaton and the smallness of $V_0$ may be related to the largeness of
such fields as, $V_0\simeq M_{\rm s}^4e^{-\chi}$.  Alternatively, one
may argue the smallness of $V_0$ is due to a nontrivial vacuum structure
which may lead to $V_0= M_{\rm s}^4e^{-S_E}$ where $S_E$ is a Euclidean
action connecting two degenerate vacua \cite{Yokoyama:2001ez} Since the
purpose of this paper is to demonstrate that it is possible to fit the
observational results, we will not pursue this issue (the problem of the
cosmological constant) any further here. }

Before proceeding to numerical calculations, let us make analytic
estimate.  From (\ref{alphahad}) and (\ref{eta}), we find
\beq
  \eta\sim 10^{-5}\alpha^2_{had}\sim 10^{-5}e^{-2\phi}
\frac{2C_g}{C_{\phi}}\,
      \lkk 
        \frac{16\pi^2}{9} A_F - \frac{2}{9} \lmk d + \frac{d_{\rm H}}{2} \rmk
        +\frac{1}{2}d_g
      \rkk^2 
 \lesssim 10^{-13}.
\eeq
Assuming the last two factors are of order of unity, we find 
$e^{-2\phi} \lesssim 10^{-7}$ which implies $\phi \gtrsim 8$.
Therefore the dilaton must have run away much beyond the Planck
scale by now.  In order to reproduce the observed time variation,
$\phi$ must change appropriately in the cosmological time scale.
Specifically, from (\ref{alphayear}) and (\ref{eq:nonalpha}) or 
(\ref{eq:SUSYalpha}), we require 
$
 {\dot{\alpha}}/{\alpha}
\sim \alpha e^{-\phi}\dot\phi \sim 10^{-65},
$
in the Planck unit with $M_{G}=1$.
Using the slow-roll equation of motion we estimate
\beq
 \dot\phi \sim \frac{V'[\phi]}{3H} \sim 
\frac{V_0d_v e^{-\phi}}{\sqrt{V_0}}\sim \sqrt{V_0}d_v e^{-\phi}
\sim 10^{-60}d_v e^{-\phi},
\eeq
in the same unit.  From these results we find that $d_v$ must take
a fairly large value, $d_v \gtrsim 10^4$. We also find a similar value
of $d_v$ from (\ref{muyear}).

The presence of nonzero $A_{\phi}$ hardly affects the evolution of the
dilaton since it is moving slowly: it only slightly facilitates 
the rolling 
of $\phi$ because the equation of motion of $\phi$ is divided by 
$C_{\phi}-A_{\phi}e^{-\phi}$. 
However, the presence of nonzero $d_g$ greatly affects the evolution 
since it induces a  {\it negative} effective potential for $\phi$: 
$V_R(\phi)\equiv -C_gd_ge^{-\phi}R/\alpha'$. $V_R$ practically vanishes
during the radiation dominated epoch 
when $R\simeq -(\alpha'/2B_g)T\simeq 0$ with 
$T$ being the trace of the energy momentum tensor,
and hence the dilaton does not move.  It becomes non-negligible during 
matter dominated epoch.  
The magnitude of $V_R$ can be larger than $V(\phi)$ and hence 
$\phi$ can {\it decrease} rather than increase \cite{extended}, quite the opposite 
to what we want. Therefore,  $d_g$ must be small enough in order to be
consistent with
the observations.

Provided these conditions are satisfied, there are a wide allowed region
in the parameter space that can account for the observed variation, so
that we only give one specific example in Fig. \ref{fig1}, where the
time variations of $\mu$ (upper graph) and $\alpha$ (lower graph) are
given.  The data of $\mu$ are taken from \cite{mpme}.  The open circles
in $\alpha$ data which are binned and averaged by redshift are taken
from \cite{webb}, crosses from \cite{SCP} and filled circles from
\cite{LCM}. We also plot the bound of $\Delta\alpha/\alpha$ from the
analysis of isotope abundances in the Oklo natural reactor operated 2Gyr
ago (corresponding to $z\simeq 0.16$). We adopt a conservative bound by
Damour and Dyson \cite{DD}.  
The bound on $\Delta\alpha/\alpha$ 
over the age of the solar system $\simeq 4.6$Gyr ($z\simeq 0.44$) from 
the comparison of the $^{187}{\rm Re}$ meteoritic measurements of the time averaged 
$^{187}{\rm Re}$ decay rate with the laboratory measurements is also included. 
We adopt a conservative bound by Fujii and Iwamoto \cite{fujii}. \footnote{
In these nulear bounds on varying $\alpha$, if quark masses are allowed to vary, 
there may be correlations between 
the effects of varying $\alpha$ and those of varying quark masses, since 
 nuclear effects depends both on electromagnetic forces and on nuclear forces. 
These effects would lead to different bounds on $\alpha$. However, we will not 
consider such a possibility here for simplicity. } 

In this example, we have considered a
non-SUSY theory and taken $d_g=0.01, C_{\phi}=A_{\phi}=1, d_v=10^4$ for
dilaton couplings to gravitational part and $A_F=0.001,d=-0.4,d_H=0.01$
for those to gauge/matter part. $C_g$ is used to normalize the
gravitational constant and set to be $0.5$, which corresponds to $M_s
\simeq M_G / \sqrt{2C_g} \simeq 2.4 \times 10^{18}$~GeV in this
example. However, one should notice that the action
(\ref{eq:action-dilaton}) is invariant under the following scale
transformation: $M_s \rightarrow a M_s, \phi \rightarrow \phi, C_g
\rightarrow C_g/a^2, C_{\phi} \rightarrow C_{\phi}/a^2, A_{\phi}
\rightarrow A_{\phi}/a^2$.  Therefore, the dynamics of the dilaton is
unchanged under the transformation.  If we take $a \simeq 1/10$, the
string scale can be lowered to the GUT scale. The initial condition of
the dilaton is $\phi=8,\dot\phi=0$ at $z=10^{10}$. As is seen in the
figure, this example can fit the observational data. Damour and Dyson
obtained $\Delta \alpha / \alpha = (-0.9 \sim 1.2) \times 10^{-7}$ at $z
\simeq 0.16$ from the Oklo mine samples \cite{DD}. In our example, $\Delta
\alpha / \alpha = -7.8 \times 10^{-8}$ at $z =0.16$, which satisfies the
Oklo bound. 
The $^{187}{\rm Re}$ bound is $|\Delta\alpha/\alpha|<1.5\times 10^{-6}$
at $z\simeq 0.44$ \cite{fujii}, but it is to be noted that the bound
depends on the way of time dependence of the $^{187}{\rm Re}$ decay rate
\cite{fujii}.\footnote{The difference between \cite{fujii} and
\cite{olive} comes from the difference in the adopted laboratory
measurements of the decay rate.  We use a more recent measurement
\cite{decayrate}.}  The bound is also satisfied in our example:
$\Delta\alpha/\alpha=-1.8\times 10^{-7}$ at $z=0.44$.
The comparison of different atomic clocks over several years
yields a bound on the present-day time variability of $\alpha$ ($\dot
\alpha/\alpha=(-0.3\pm 2.0) \times 10^{-15}~{\rm yr}^{-1}$ \cite{Peik}),
which is also satisfied in our example.  We also note that this example
reproduces the cosmic expansion law of the present universe correctly
since the dilaton potential plays the role of the dark energy with
$\Omega_{d}=0.74$ and the equation-of-state parameter $w\simeq -0.91$.

Although the allowed region in the parameter space is large we mention
that there is a tension to account for both the smallness of $\eta$ and
the observed values of $\dot{\alpha}/\alpha$ and $\dot\mu/\mu$
simultaneously, for the latter requires a relatively large $\dot\phi$
sourced by a large $d_v$.  In the present example, we find
$\gamma-1\simeq -2.9\times 10^{-9}$ and $\eta\simeq 7.6\times 10^{-14}$.

Finally we comment on the time variation of the gravitational 
constant $G$. 
It is given by $\dot G/G=2\alpha_{had}\dot\alpha_{had}/(1+\alpha_{had}^2)<0$. 
However, its magnitude  is very small due to the 
smallness of $\alpha_{had}$: 
$|\dot G/G|\sim -2\alpha_{had}|\dot\alpha/24\alpha^2|\sim 10^{-18}{\rm yr}^{-1}$ 
and safely satisfies the present experimental constraint \cite{gdot}:
$\dot G/G =(4\pm 9)\times 10^{-13}{\rm yr}^{-1}$.  
The tests of the weak equivalence principle put severe limits on dilaton models.  
More precise experiments of the weak equivalence principle 
could discover its violation by a runaway dilaton.

\section{Conclusion}

Motivated by the observational evidence that indicates nontrivial
time evolution of the fundamental constants such as the 
proton-electron mass ratio and the fine structure constant, 
we have theoretically calculated
their time variations based on the standard particle-physics model
and its supersymmetric extension with the help of the renormalization
group approach.  We have employed several simple assumptions such
as the universality of the time variation of the masses of the 
fundamental fermions as a first step.
We have applied our formalism to a specific scenario of the 
runaway dilaton and found that we can account for the observed
time variation with some natural choice of model parameters.

We have also found that there is some tension between the observed
magnitude of the time variation and the experimental constraint imposed
by the verification of the equivalence principle.  Indeed we typically
find $\eta$ to be larger than $10^{-13}$ unless we adopt a sufficiently
small value of $A_F$. One can regard this feature of our model as a
prediction, that is, if one performs a more precise experiment on the
equivalence principle to measure $\eta$ with a higher accuracy, one
would be able to discover its violation by a runaway dilaton.

\section*{Acknowledgments}

This work was partially  supported  by JSPS Grant-in-Aid 
for Scientific 
Research Nos.~16340076(JY), 17204018(TC), 17540251(TK) and 18740157(MY).
Also, T.~C.\/ is supported in part by Nihon University. 
T.~K.\/ is also supported in part by 
the Grant-in-Aid for
the 21st Century COE ``The Center for Diversity and
Universality in Physics'' from the Ministry of Education, Culture,
Sports, Science and Technology of Japan.
M.Y. is supported in part by the project of the Research Institute of
Aoyama Gakuin University.

\end{document}